\documentclass[aps,pra,floatfix,amsmath,ams,preprint,superscriptaddress,showpacs]{revtex4-1}
\usepackage{graphicx}
\usepackage{dsfont}
\begin{document}

\title{On the Kolmogorov-Sinai entropy of many-body Hamiltonian systems}

\author{Arul Lakshminarayan }
\affiliation{Department of Physics, Indian Institute of Technology Madras, 
Chennai, 600036, India}

\author{Steven Tomsovic }
\affiliation{Department of Physics, Indian Institute of Technology Madras, 
Chennai, 600036, India}
\affiliation{Department of Physics and Astronomy, Washington State University, Pullman, Washington 99164-2814 USA\footnote{permanent address}}


\begin{abstract}
The Kolmogorov-Sinai (K-S) entropy  is a central measure of complexity and chaos.  Its calculation for many-body systems is an interesting and important challenge.  In this paper, the evaluation is formulated by considering $N$-dimensional symplectic maps and deriving a transfer matrix formalism for the stability problem.  This approach makes explicit a duality relation that is exactly analogous to one found in a generalized Anderson tight-binding model, and leads to a formally exact expression for the finite-time K-S entropy.  Within this formalism there is a hierarchy of approximations, the final one being a diagonal approximation that only makes use of instantaneous Hessians of the potential to find the K-S entropy.  By way of a non-trivial illustration, the K-S entropy of $N$ identically coupled kicked rotors (standard maps) is investigated.  The validity of the various approximations with kicking strength, particle number, and time are elucidated.  An analytic formula for the K-S entropy within the diagonal approximation is derived and its range of validity is also explored.  
 
\end{abstract}
\pacs{5.45.Jn,5.45.Ra,5.45.Pq,2.70.Hm}
\maketitle


\newcommand{\newc}{\newcommand}
\newc{\beq}{\begin{equation}}
\newc{\eeq}{\end{equation}}
\newc{\kt}{\rangle}
\newc{\br}{\langle}
\newc{\beqa}{\begin{eqnarray}}
\newc{\eeqa}{\end{eqnarray}}
\newc{\pr}{\prime}
\newc{\longra}{\longrightarrow}
\newc{\ot}{\otimes}
\newc{\rarrow}{\rightarrow}
\newc{\h}{\hat}
\newc{\bom}{\boldmath}
\newc{\btd}{\bigtriangledown}
\newc{\al}{\alpha}
\newc{\be}{\beta}
\newc{\ld}{\lambda}
\newc{\sg}{\sigma}
\newc{\p}{\psi}
\newc{\eps}{\epsilon}
\newc{\om}{\omega}
\newc{\mb}{\mbox}
\newc{\tm}{\times}
\newc{\hu}{\hat{u}}
\newc{\hv}{\hat{v}}
\newc{\hk}{\hat{K}}
\newc{\ra}{\rightarrow}
\newc{\non}{\nonumber}
\newc{\ul}{\underline}
\newc{\hs}{\hspace}
\newc{\longla}{\longleftarrow}
\newc{\ts}{\textstyle}
\newc{\f}{\frac}
\newc{\df}{\dfrac}
\newc{\ovl}{\overline}
\newc{\bc}{\begin{center}}
\newc{\ec}{\end{center}}
\newc{\dg}{\dagger}
\newc{\prh}{\mbox{PR}_H}
\newc{\prq}{\mbox{PR}_q}
\newc{\tr}{\mbox{tr}}
\newc{\pd}{\partial}
\newc{\qv}{\vec{q}}
\newc{\pv}{\vec{p}}
\newc{\dqv}{\delta\vec{q}}
\newc{\dpv}{\delta\vec{p}}
\newc{\mbq}{\mathbf{q}}
\newc{\mbqp}{\mathbf{q'}}
\newc{\mbpp}{\mathbf{p'}}
\newc{\mbp}{\mathbf{p}}
\newc{\mbn}{\mathbf{\nabla}}
\newc{\dmbq}{\delta \mbq}
\newc{\dmbp}{\delta \mbp}
\newc{\T}{\mathsf{T}}
\newc{\J}{\mathsf{J}}
\newc{\sfL}{\mathsf{L}}
\newc{\C}{\mathsf{C}}
\newc{\B}{\mathsf{M}}
\newc{\V}{\mathsf{V}}

\section{Introduction}

The Kolmogorv-Sinai (K-S) entropy is a widely used measure of chaos in dynamical systems~\cite{Lichtenberg83,Ott02} and indicates the exponential rate at which information is produced due to the complex dynamics.  It is also the exponential rate at which certain $N$-dimensional phase space differential surfaces expand, where $N$ is the number of degrees of freedom.  It is positive only for chaotic systems and vanishes otherwise. According to Pesin's theorem~\cite{Eckmann85}, in a closed fully chaotic system it is also the sum of all positive Lyapunov exponents,  which measure the local exponential divergence of phase-space trajectories.  In an open or scattering system, the difference between the sum of the positive Lyapunov exponents and the K-S entropy may be related to  transport properties like diffusion coefficients~\cite{Gaspard90}, thus providing a link between the macroscopic and microscopic dynamics.  Given its central importance, it has not only been calculated in low-dimensional and model systems, but also in many-body systems  such as dilute hard sphere/disk gases~\cite{Beijeren97, Dorfman98,Astrid05} and dilute wet granular gases~\cite{Fingerle07}.  It transplants the notion of entropy, originally introduced as a measure of disorder in statistical mechanics, into the context of low-dimensional  deterministic dynamical systems. 

The finite-time K-S entropy turns out to be the natural measure that appears in semiclassical mechanics and provides, along with the principle of uniformity of periodic orbits and periodic-orbit information (i.e. actions, periods, and Maslov indices), the route to inferring quantum spectra from classical dynamics via the Gutzwiller trace formula~\cite{Gutzwiller71}.  It appears to have important connections to the rate of entropy or entanglement production in strongly chaotic, but weakly coupled systems, the rate being proportional to the K-S entropy~\cite{Miller99}.  Finite-time Lyapunov exponents have also proven useful in many contexts and were recently used for instance as indicators of the presence of very small regular islands in an otherwise chaotic phase space~\cite{Tomsovic07,Vallejo08,Manchein09}.  In high-dimensional systems, it seems more natural to study a finite-time K-S entropy and look at the fluctuations of this quantity in an ensemble of trajectories as a potential measure that can indicate non-ergodicity.

This paper addresses the issue of calculating the K-S entropy from these motivations and restricts attention to Hamiltonian systems in the form of symplectic maps~\cite{Lichtenberg83}.  In particular, the connection to a transfer-matrix formulation is made explicit, wherein a (transformed) Jacobian plays this role, and the problem is  formulated in terms of the spectra of block tridiagonal Hermitian matrices.  A general duality relation emerges that has an exact mapping to previous work on Anderson models of transport~\cite{Molinari98,Molinari03}.  A generalization of the Thouless relation gives an exact formula for the finite time K-S entropy in terms of an average over a phase at the boundary.  Although the connections between Lyapunov exponents of transfer matrices and localization length through transfer matrices have been well known for quite some time \cite{Pichard81}, to our knowledge the generalization to the K-S entropy of classical symplectic maps is absent.  Nevertheless, the ingredients from the localization and generalized transfer matrix literature are indeed there \cite{Molinari98}.
   
In this paper the focus is on the the infinite time limit (asymptotic) entropy.  The formalism is used to derive a simply calculable upper bound for the K-S entropy that  becomes better the stronger the chaos as well as with the number of particles in the system.  Furthermore, it is possible to argue for a series of increasingly restrictive approximations culminating in the  ``diagonal approximation" of the K-S entropy that gives estimates often much better than the upper bound in regimes of fully developed chaos.  The diagonal approximation was discussed in some detail recently in the context of two-dimensional maps via the standard map~\cite{Tomsovic07}.  A generalization to higher dimensions is provided herein. The treatment, though restricted to symplectic maps or Poincar\'e sections of Hamiltonian flows, can conceivably be usefully extended to continuous flows. 

Both the diagonal approximation and the upper-bound are defined in terms of spectra of instantaneous Hessians of a potential (through the Hessian of the action).  Studying stability problems using the Hessians of potentials to calculate quantities of a similar nature as the K-S entropy is not new.  They appear naturally in stability problems as  part of a curvature.   For instance, they appear in using Riemannian geometry to study stability, in particular to find the largest Lyapunov exponent~\cite{Casetti96}. Also, a Hamilton-Jacobi formulation of the K-S entropy~\cite{Partovi02} uses similar quantities. The approach presented in this paper, though, is different as it is based on generating functions and  generalized transfer matrices.  It provides routes for several approximations and bounds in both the chaotic and moderately chaotic regimes.

The diagonal approximation is applied to the K-S entropy of a system of $N$ strongly interacting kicked rotors, which may be considered to be a kicked version of the well-studied Hamiltonian mean field model~\cite{Antoni95}, and which has also been studied for sometime~\cite{Konishi92}.  That interest continues, for example, Ref.~\cite{Woellner09} is related to the present work.  They calculate the K-S entropy numerically, among other results.  An analytical expression for the K-S entropy is derived in this paper by finding the density of states of the potential Hessian.  This is perhaps the first analytical evaluation of the K-S entropy in a nontrivial model of a many-body chaotic system of this kind, whereas for the two-dimensional standard map (kicked rotor) the Chirikov estimate of the Lyapunov exponent or the K-S entropy has long been known~\cite{Chirikov79}. 

\section{From transfer matrices to the K-S entropy}

Consider an $N$-degree-of-freedom system with the phase space coordinates $(\mathbf{q},\mathbf{p})$. The discrete time evolution indicated by the subscripts is derived from a generating function $F(\mbq,\mbqp)$ of old ($\mbq)$ and new ($\mbqp)$ coordinates,
\beq
\mbp= -\mbn_{\mbq} F(\mbq,\mbqp),\;\; \mbpp= \mbn_{\mbqp} F(\mbq,\mbqp)\ .
\eeq
This follows from the stationarity of the action $W(\mbq_0,\ldots, \mbq_t) = \sum_{i=0}^{t-1} F(\mbq_{i},\mbq_{ i+1})$ at the interior points ($1 \le i < t)$ of  trajectories of length $t+1$.  The action derivatives at the edges of the trajectory are non-zero, giving the momenta at these points.  The analogy with transfer matrices becomes more complete ahead with the introduction of the auxilliary generating function 
\beq
\tilde{F}(\mbq,\mbqp)=F(\mbq,\mbqp)-E |\mbq|^2/2
\eeq
where $E$ is a real number.  This is equivalent to adding an isotropic harmonic oscillator of frequency $\sqrt{E}$ to the potential, whose frequency set to zero recovers the original dynamics.  Denoting the discrete time as subscripts, the Jacobian $\J_i$ matrix propagates phase space variations, or is the tangent map:
\beq
\J_{i} \left( \begin{array}{c} \dmbq_i \\ \dmbp_{i} \end{array} \right) = \left( \begin{array}{c} \dmbq_{i+1} \\ \dmbp_{i+1} \end{array} \right).
\eeq
Define the transfer matrix $\T_i$ as 
\beq
\mathsf{T}_{i} \left( \begin{array}{c} \dmbq_i \\ \dmbq_{i-1} \end{array} \right) = \left( \begin{array}{c} \dmbq_{i+1} \\ \dmbq_{i} \end{array} \right)
\eeq
and let $\sfL_i$ be the matrix such that
\beq
\mathsf{L}_{i} \left( \begin{array}{c} \dmbq_i \\ \dmbp_{i} \end{array} \right) = \left( \begin{array}{c} \dmbq_{i} \\ \dmbq_{i-1} \end{array} \right)\ .
\eeq
The following relation then holds
\beq
\T_i = \sfL_{i+1}\, \J_i \sfL_i^{-1}, \; \; \J_i= \sfL_{i+1}^{-1} \T_i \sfL_i.
\eeq

Defining the matrices $\C_i$ and $\B_{i-1}$ as
\beq
(\C_i)_{kl}= \df{\pd^2 F(\mbq_{i-1},\mbq_i)}{\pd(\mbq_i)_k \pd(\mbq_i)_{l}},\;\; 
(\B_{i-1})_{kl}= -\df{\pd^2 F(\mbq_{i-1},\mbq_i)}{\pd(\mbq_i)_k \pd(\mbq_{i-1})_{l}}
\eeq 
gives
\beq
\dmbp_i=\C_i \dmbq_i -  \B_{i-1} \dmbq_{i-1}\ ,
\eeq
where
\beq
\sfL_i = \left( \begin{array}{cc} I &0 \\ \B_{i-1}^{-1} \C_i& -\B_{i-1}^{-1} \end{array} \right), \;\;
\sfL_i^{-1} = \left( \begin{array}{cc} I &0 \\ \C_i& -\B_{i-1} \end{array} \right).
\eeq
Note that while $\C_i$ is a symmetric matrix, $\B_i$ may be, but generally is not.  In the following, it is assumed that $\B_i$ is a positive semidefinite matrix, which implies that the symplectic map generated by it is a twist map.  The Jacobian and transfer matrices can be written in terms of the matrices introduced above with the addition of $\tilde{\C}$ defined as 
\beq
(\tilde{\C}_i)_{kl} =  \df{\pd^2 F(\mbq_{i},\mbq_{i+1})}{\pd(\mbq_i)_k \pd(\mbq_i)_{l}}.
\eeq
Denoting the dependence on the real parameter $E$ as $\J_i(E)$ the Jacobian is
\beq
\J_i(E)=\left( \begin{array}{cc} -(\B_i^T)^{-1}(E-\tilde{\C}_i) & (\B_i^T)^{-1}\\
-\C_{i+1}(\B_i^T)^{-1}(E-\tilde{\C}_i) - \B_i & \C_{i+1}(\B_i^T)^{-1}
\end{array} \right)
\eeq
and the transfer matrix is 
\beq
\T_i(E)=\left( \begin{array}{cc} (-\B_i^T)^{-1}(E-\tilde{\C}_i-\C_i) & -(\B_i^T)^{-1}\B_{i-1} \\
I & 0
\end{array} \right)
\eeq

The Jacobian and transfer matrices over a time $t$ are the products $\J(E)=\J_t(E) \J_{t-1}(E) \cdots \J_1(E)$ and $\T(E)=\T_t(E) \cdots \T_1(E)$, respectively.  By construction, $\J(E)$ is a symplectic matrix, but the transfer matrix generally is not.  To relate the spectral properties of $\J(E)$, which is of interest, to those of transfer matrices, a closely related matrix 
\beq
\tilde{\T}(E)=\sfL_1 \sfL_{t+1}^{-1} \T(E)
\eeq
is introduced.  This matrix must be isospectral with $\J(E)$ since the pair are related by a similarity transformation, $\J(E)=\sfL_1^{-1} \tilde{\T}(E)\sfL_1$.  It follows also that $\tilde{\T}(E)$ is symplectic, 
\beq
\tilde{\T}^T(E) \Sigma \tilde{\T}(E) = \Sigma,\;\; \Sigma= \left( \begin{array}{cc}0 & \B_0\\ -\B_0^T&0 \end{array} \right)\ .
\eeq
Therefore the characteristic polynomial of $\J(E)$, $P(\ld)= \det[\J(E)-\ld I]$ is identical to that of $\tilde{\T}(E)$, and is  reflexive, {\it i.e.} $P(\ld)= \ld^{2N} P(1/\ld)$.  Note that in the cases of the central orbit being {\it periodic} of period $t$ or the $\sfL_i$ independent of time, $\tilde{\T}(E)$ coincides with the transfer matrix $\T(E)$.  The latter is the case for many multi-dimensional symplectic maps considered in the literature so far. 

The advantage of the transfer matrix approach, rather than working directly with the quantity of interest, i.e.~the Jacobian, is that it immediately motivates the
construction of a ``Hamiltonian'', not to be confused with the one which may be generating the dynamics.  Before constructing this fictitious Hamiltonian, the ``boundary'' conditions need to be ascertained.  Let $(\dmbq_1,\dmbq_0)^T$ be an eigenvector of $\tilde{T}(E)$ with eigenvalue $\ld$, then 
\beq
\sfL_1 \sfL_{t+1}^{-1} \left( \begin{array}{c} \dmbq_{t+1}\\ \dmbq_t \end{array} \right)=
\ld \, \left( \begin{array}{c} \dmbq_{1}\\ \dmbq_0 \end{array} \right)
\eeq
which implies that
\beqa
\dmbq_{t+1}&=&\ld \, \dmbq_1 \nonumber \\
\dmbq_0&=&-\B_0^{-1}(-\C_1+\C_{t+1}) \dmbq_1 + \df{1}{\ld} \B_0^{-1} \B_t \dmbq_t
\eeqa
Again, if the orbit was periodic with period $t$ or if $L_i$ was independent of time, the conditions would be simpler.  Using the transfer matrix, a vector of dimension $Nt$ can be built out of the initial variation that is an eigenvector of $\tilde{T}(E)$: $\psi_{\ld,E}=(\dmbq_1,\cdots, \dmbq_t)^T$.  Using the above boundary conditions it follows that $Nt$-dimensional vector satisfies the eigenvalue equation 
\beq
{\cal H}(\ld) \psi_{\ld,E}= E \psi_{\ld,E}
\eeq
where ${\cal H}(\ld)$ is the $Nt$ dimensional matrix:
\beq
\label{ham}
\left( \begin{array}{cccccc}
{\tilde \C}_1+\C_{t+1} & -\B_1^T & 0 & \cdots & 0 & -\df{1}{\ld}\B_t\\
-\B_1 & \tilde{\C}_2+\C_2 & -\B_2^T &0& \cdots & 0\\
0 &-\B_2 & \tilde{\C}_3 +\C_3& -\B^T_3 & \cdots & 0 \\
\vdots&\vdots&\vdots&\vdots& \vdots& \vdots \\
-\lambda \B_t^T&0 &0 &\cdots& -\B_{t-1}&\tilde{\C}_t + \C_t\end{array}
\right). 
\eeq
Thus there is a duality in the sense that if and only if $\tilde{\T}(E)$ has an eigenvalue $\ld$, ${\cal H}(\ld)$ has the eigenvalue $E$. The ``Hamiltonian" above is the Hessian of the action when $\ld=1$.

The matrix ${\cal H}(\ld)$ is Hermitian in the case $\ld$ lies on the unit circle. It is a block-tridiagonal matrix with corner blocks.  When $N=1$, such a formulation for periodic orbits has been known for some time and used by Bountis, Helleman~\cite{Bountis81} and Greene~\cite{Greene87} to study stability of orbits in the Chirikov-Taylor standard map. For $N>1$ an extension was discussed, again for periodic orbits, by Kook and Meiss~\cite{Kook89}.  In all these cases, $E=0$ was the relevant parameter.  The extension here is valid for arbitrary orbits and includes a parameter that makes the analogy with the literature on generalized tight-binding models direct.  In this context, the works of Molinari~\cite{Molinari98,Molinari03} discuss the extension beyond one-dimension and applications include multichannel scattering in mesoscopic systems.  There, $E$ is the energy and the eigenvalues of the transfer matrix relate to localization.  The relation of Eq.~(\ref{ham}) follows by considering how variations evolve under a Newtonian formulation of dynamics in terms of second difference operators, however the above method explicates the structure of the transfer matrix, and its relationship to the Jacobian.  Classical mechanics is naturally equipped with these structures.

There is a simple relationship between the characteristic polynomial of ${\cal H}(\ld)$ and $P(\ld)$, the characteristic polynomial of $\J(E)$ and $\tilde{T}(E)$. This is essentially the ``duality" relation that was discussed in the context of transfer matrices by Molinari and is a generalization of that stated by Kook and Meiss for periodic orbits \cite{Kook89}. The simultaneous vanishing of the polynomials implies their proportionality: $P(\ld) \sim \det[{\cal H}(\ld)-E]$. $P(\ld)$ is a polynomial in $\ld$ of order $2N$, with the leading term being $\ld^{2N}$, and $ \det[{\cal H}(\ld)-E]$ runs from $\lambda^N$ to $\lambda^{-N}$. Thus the coefficient of $\ld^N$ in $\det[{\cal H}(\ld)-E]$ provides the proportionality constant. Using the basic definition of the determinant as a signed  sum of products of matrix elements whose row and column indices are permutations, it is possible to conclude that this is $(-1)^{N} \det(\B_1\cdots \B_t)$ \cite{Molinari08}.  This gives the duality relation stated in terms of the original Jacobian matrix polynomial as:
\beq
P(\lambda)=\det(\J(E)-\ld I) = (-\ld)^{N} \det(\B_1\cdots \B_t)^{-1} \det[{\cal H}(\ld)-E I].
\eeq
It is consistent with the relations mentioned for periodic orbits in \cite{Kook89}  as well as in the work on transfer matrices. 

There are several consequences of this duality relation that can be derived, and are potentially useful in estimating entropies of dynamical systems.  First consider the case $E=0$, which corresponds to the generating function of interest.  Let the set of eigenvalues  of $\J \equiv \J(0)$  be $\{\exp(t \gamma_k+i t \phi_k)\}$.  From the reflexive property it follows that the reciprocal is also an eigenvalue and from the reality of the polynomial (even if $E\ne 0$ it is always assumed real here) that the complex conjugate is also an eigenvalue.   Thus the eigenvalues either come in pairs, in the case $\gamma_k$ or $\phi_k$ is zero,  or in quartets.  If the orbits are periodic, these different cases correspond to elliptic, hyperbolic and loxodromic orbits respectively.  The numbers $\{\gamma_k\}$ are important and indicate the exponential instability, they are finite-time stability exponents and are closely allied to the well-studied finite-time Lyapunov exponents. In the infinite time limit for generic systems they comprise the complete spectrum of Lyapunov exponents.  The sum of the finite-time positive elements may therefore be considered more or less a finite-time K-S entropy ($h_t$). 

An exact formula for this finite-time K-S entropy $h_t$ follows by applying Jensen's formula~\cite{Conway} to the analytic function $P(\lambda)$, where $\lambda$ is now considered as a complex variable.  Jensen's formula is a generalization of the Mean Value Principle of harmonic functions, and although $\log[|P(\lambda)|]$ is not a harmonic function on the unit circle centered at the origin due to the zeros of $P(\lambda)$ in the interior, these zeros can be removed in a suitably redefined function. That results in: 
\begin{eqnarray}
h_t=\sum_{\gamma_k>0} \gamma_k &=&  \int_{0}^{2 \pi} h_t(\theta) \df{d\theta}{2 \pi} \nonumber \\
&=& -\df{1}{t}\sum_{i=1}^{t}\ln[\det(\B_i)] + \df{1}{t}\int_{0}^{2 \pi} \ln(|\det[{\cal H}(e^{i \theta})]|) \df{d\theta}{2 \pi}.
\end{eqnarray}
From this exact relation for $h_t$ comes the first level of approximation.  Under some circumstances, such as long times and/or chaos that is not too weak, the $\theta$-dependence of $h_t(\theta)$ becomes negligible.  Without going into details just yet (to be discussed ahead), for the many-body kicked rotors there are very interesting, but not fully understood, regimes of behavior which emerge on this point.  In particular and unexpectedly, for some initial conditions, values of interaction strength, times,  and to the accuracy of the numerical calculations done,
\beq
\tilde h_t(0)= \int_{0}^{2 \pi} h_t(\theta) \df{d\theta}{2 \pi} 
\eeq
where $\tilde h_t(0)$ indicates that the time transient contribution due to a center of mass coordinate is first subtracted from $h_t(0)$.  In those cases, $h_t=\tilde h_t(0)$ effectively is not an approximation and it is possible to study finite-time K-S entropy fluctuations via the simpler quantity $\tilde h_t(0)$.  It could be useful for finding non-ergodic dynamics following the methods of Refs.~\cite{Tomsovic07,Vallejo08,Manchein09}, but is outside the scope of this paper.

In the limit of large times, $h_t$ reaches its asymptotic value, which is the K-S entropy
\beq
\label{defhksh}
h_{KS} = \lim_{t \rarrow \infty} \df{1}{t} \ln|\det(\J-I)|=\lim_{t \rarrow \infty} -\df{1}{t}\sum_{i=1}^{t}\ln[\det(\B_i)] + \df{1}{t} \ln(|\det[{\cal H}(1)]|)
\eeq
($\theta$ averaging no longer needed).  The quantity defined herein, $|\det(\J-I)|$, appears in many considerations of a semiclassical kind and the study of its behaviors and fluctuations are of considerable interest~\cite{Gutzwiller90,Elton10}.  If the normalized density of eigenvalues of ${\cal H}(1)$ is denoted as $\rho_H(\mu)$, the entropy can be written as 
\beq
\label{hksh}
h_{KS}=-\left \langle \ln(\det(\B_i)) \right \rangle + N \int \ln(|\mu|) \rho_H(\mu)\, d \mu.
\eeq
 Although the spectral problem seems to have been compounded by now focussing attention on a $Nt$-dimensional matrix ($t\rightarrow\infty)$ instead of products of $2N$ dimensional ones, in fact the eigenvalues of the large matrix are usually bounded, whereas those of the product exponentially increase.  
 
 A second level of approximation suggests itself here and is denoted the banded approximation.  As the spectrum depends little on the nature of the corner blocks, they may as well be neglected entirely.  In that case, $\cal H$ is truly banded and its determinant and spectrum can replace that of ${\cal H}(1)$ in Eqs.~(\ref{defhksh},\ref{hksh}).  This approximation can be extremely useful numerically as the diagonalization can be performed much more efficiently.  Very little is lost in making this approximation and it is expected to be valid even well into the weakly chaotic regime.  It may yet be possible to study finite-time fluctuations at this level of approximation.   
  
\section{An upper bound, the diagonal approximation, and an estimate}

The following considerations are for a specialized class of generating functions, mechanical-type, that are still general enough to contain many of the Hamiltonians studied in the literature on coupled maps. The idea is to focus on the essential details and facilitate interpretation, as the generalizations are straightforward.
Let
\beq
F(\mbq_i,\mbq_{i+1})=\f{1}{2} (\mbq_i-\mbq_{i+1})\cdot \B \cdot  (\mbq_i-\mbq_{i+1}) -V(\mbq)
\eeq
where $\B$ is a constant symmetric positive semidefinite matrix. Then the symplectic map is
\beqa
\mbq_{i+1}&=\mbq_i+\B^{-1} \mbp_{i+1}\\
\mbp_{i+1}&=\mbp_i-\nabla V(\mbq_i).
\eeqa
In this case the $\sfL_i$ matrices are independent of time $i$, and the instantaneous transfer
matrix $\T_i$ is a similarity transform of the Jacobian matrix $\J_i$.  The ``Hamiltonian" is 
\beq
{\cal H}(1)=
\left( \begin{array}{cccccc}
-\V''_1+2\B & -\B & 0 & \cdots & 0 & -\B\\
-\B & -\V''_1+2\B & -\B &0& \cdots & 0\\
0 &-\B & -\V''_2+2\B & -\B & \cdots & 0 \\
\vdots&\vdots&\vdots&\vdots& \vdots& \vdots \\
-\B&0 &0 &\cdots& -\B&-\V''_{t}+2\B \end{array}
\right) 
\eeq
where $\V''_i$ is the instantaneous Hessian matrix of the potential.  Thus in this case the first block of the matrix remains as if the orbit were periodic. 

The matrix ${\cal H}(1)$ although Hermitian is not positive semidefinite.  Consider the diagonal $N\times N$ blocks of the positive definite matrix ${\cal H}(1)^2$.  These are $2\B^2+(2\B-\V''_i)^2$, which are themselves clearly positive semidefinite.  A generalization of the well-known Hadamard determinant inequality may be applied. If $A$ is a positive semidefinite matrix whose diagonals are partitioned into $t$, $N-$dimensional square matrices $P_i$ that are themselves positive semidefinite, then 
$\det(A) \le \Pi_{i=1}^t  \det(P_i)$.  In the instance that $N=1$ and the $P_i$ are simply the diagonal elements of $A$, this is the usual Hadamard inequality.  The inequality generalization provides a better upper bound and may be compared with other generalizations of the inequality such as Fischer's~\cite{Horn85}.  Applying this inequality with $A={\cal H}(1)^2$ results in an upper-bound  $\ovl{S}$ for the K-S entropy
\beqa
h_{KS} \le \ovl{h}_{KS}&=&\lim_{t \rarrow \infty} \df{1}{2t}\sum_{i=1}^t \ln\left(\det\left[2I+(2I-\B^{-1/2}\V''_i \B^{-1/2})^2\right]\right) \nonumber\\
&=& \left \br \df{1}{2}\ln\left(\det\left[2I+(2I-\B^{-1/2}\V'' \B^{-1/2})^2\right]\right)\right \kt=\df{N}{2} \int \ln(2 +\mu^2) \rho_V(\mu)\, d \mu\nonumber\\
\eeqa
where $\B^{-1/2}$ is the unique positive semidefinite square-root of the matrix $\B$. The product of the mass matrix with the potential Hessian is written such that the resulting matrix remains symmetric, however this is not necessary and the form $\B^{-1} \V''$
maybe used instead. The  angular brackets indicate replacing the time-average by a space-average, assuming ergodicity as in the case  of chaotic dynamics. The final equality is written in terms of $\rho_V(\mu)$, the normalized density of eigenvalues of the $N$ dimensional symmetric matrices $2I-\B^{-1/2}\V''_i \B^{-1/2}$.  An ensemble is formed out of the instantaneous potential Hessians to construct such a density.  For many systems this is an easily calculable upper-bound depending on the Hessian of the potential. In terms of an interpretation, the local instantaneous Hessians provide an equivalent set of harmonic oscillators and the instability is bounded by this system.  The larger the potential curvature is, the larger the upper-bound can be.  Surprisingly, the upper-bound can almost be reached, and it gets better with increased chaos and/or particle number. 

As the upper-bound is dominated by the diagonal blocks with the Hessian of the potential it may be expected in strongly chaotic cases that  just the diagonal blocks provide an approximation to the whole matrix. This constitutes the ``diagonal approximation" which can be written as the assertion that:
\beq
\label{diagonal}
h_{KS} \approx h_d= \lim_{t \rarrow \infty}\df{1}{t} \sum_{i=1}^{t} \ln \left|\det\left(2I -\B^{-1/2}\V''_i \B^{-1/2} \right) \right |=
 N \int \ln|\mu| \rho_V(\mu) \, d\mu.
\eeq
 In fact for the diagonal approximation to be valid it must turn out that $\rho_H(\mu) \approx \rho_V(\mu)$, thus ignoring the different ways the mass matrices enter the definitions. 

A detailed study of this approximation was made for the case of the Chirikov-Taylor standard map in~\cite{Tomsovic07}.  Note that this is a ``local" approximation in the sense that it is built out of adding contributions at each instant of time. The above derivation does not estimate errors and may be regarded at this stage as heuristic.  However, recall that for the one-dimensional case it is possible to write an expansion of the determinant and estimate errors.  In fact it is worth pointing out  that for the one-dimensional standard map with $V(q)=-(K/4 \pi^2) \cos(2 \pi q)$, the upper bound is given by
\beq
\ovl{h}_{KS}= \df{1}{2}\int_0^{1} \ln\left(2+(2-K \cos(2 \pi q))^2\right)\, dq = \ln\frac{|K|}{2}+\frac{1}{|K|} +O(K^{-2}),
\eeq 
which is close to the widely used estimate of Chirikov \cite{Chirikov79} that coincides
with the diagonal approximation:
\beq
h_d = \int_0^{1} \ln\left|2-K \cos(2 \pi q)\right| \, dq = \ln\frac{|K|}{2}\ .
\eeq
A more detailed and careful analysis of this case gave 
\beq
\label{chirikov}
h_{KS}=\ln\frac{|K|}{2}+\frac{1}{K^2-4} + O(K^{-6})
\eeq
thus showing that the diagonal approximation is closer to the ``exact" entropy (in the 
one-dimensional case this is the same as the Lyapunov or stability exponent) than
is the upper-bound.  Nevertheless, both the approximation and the bound are indeed close to the actual value.

Consider for a moment that under the action of completely chaotic dynamics, the entries into the inverse mass matrix product with the Hessian become something like random entries.   The determinant  in Eq.~(\ref{diagonal}) could be treated as a random variable with normalized probability density $\rho_V(\mu)$ and a variance $\sigma^2$.  In that case
\beq
\label{diagapprox}
\f{h_{KS}}{N}=  \int_{-\infty}^\infty \ln \left| \mu  \right| \rho_V(\mu) d \mu = \ln \sigma + \int_{-\infty}^\infty \ln \left| y \right| \tilde{\rho}(y) dy,
\eeq
where $y=\mu/\sigma$ and $\tilde{\rho}(y)$ has only weak dependence on particle number or interaction strength (assuming the system is chaotic).  Thus, the result is the logarithm of the width of the density plus a remaining integral that is nearly a constant.  In the case of a Gaussian density for $\rho_V(\mu)$, the constant is $-(\gamma +\ln 2)/2$, where $\gamma$ is the Euler-Mascheroni constant.  In the case of a lognormal density, the constant is unity.  If the distribution is not exactly zero-centered in its mean, then corrections emerge ordered in powers of the ratio of the mean to standard deviation.  For example, the positive-off-centered Gaussian density gives
\beq
\label{gaussianentropy}
\begin{split}
\f{h_{KS}}{N} &= {1\over \sqrt{2\pi \sigma^2}} \int_{-\infty}^\infty {\rm d}\mu \ln |\mu| \exp\left( - {(\mu-\mu_0)^2 \over 2 \sigma^2} \right) \\
&= \ln \sigma -{\gamma + \ln 2 \over 2} - \sum_{k=1}^{\infty}  \f{(-1)^k}{2k (2k-1)!!} \left( \f{\mu_0}{\sigma}\right)^{2k} \\
&\sim \ln \mu_0 - \sum_{k=1} \frac{(2k-1)!! }{2k} \left(\frac{\sigma}{\mu_0}\right)^{2k}\ {\rm \quad (asymptotic\ series)} \ ,
\end{split}
\eeq
a formula that is used ahead with minor modifications.  Two forms are given for the result of the integral.  The first is the convergent series, whereas the second is the asymptotic series.  The latter is far more useful due to slow convergence of the former if the width is much smaller than the mean.  Finally note that some significant  information is lost in making the diagonal approximation.  There is little reason to expect it to work well enough numerically to study finite time approximations just by limiting the number of consecutive instantaneous Hessians (the analytical expression is already fully ergodically averaged and without fluctuation information) and it cannot be expected to hold in the weakly chaotic regime as well as the banded approximation.

\section{K-S entropy of a coupled map lattice }
\label{identical}

Consider $N$ identical particles each of whose configuration space is a circle and interacting via two-body forces that depend only on the angular distance $|q_i-q_j|$; in this section the notation is changed such that subscripts label particles rather than time.  Assume the mass matrix is diagonal with unit elements.  The potential can be represented as $V = \sum_{i<j } v(|q_i-q_j|)$ and there is a center of mass degree of freedom that can be removed from the calculation for any number of particles.  Conveniently, the two particle case reduces to the one particle case with a renormalized interaction strength and can be used as a check.  Before removing the center of mass, the Hessian matrix has the property that sums of row or column elements vanishes.  

In order to work through a concrete system in detail, a system of coupled
Chirikov type maps is introduced that has been studied for some time and has many interesting features~\cite{Konishi92}.  It can also be considered as the kicked version
of the well-studied Hamiltonian mean field model~\cite{Antoni95}.  The Hamiltonian is
\beq
H= \sum_{i=1}^N\f{p_i^2}{2}+ \f{K}{4 \pi^2 \sqrt{N}}\sum_{i<j} \cos 2\pi (q_i-q_j) \, \sum_{n=-\infty}^{\infty}\delta(t-n),
\eeq
and the resulting map connecting variables just before subsequent kicks is 
\beq
\begin{split}
p_i'&= p_i +\f{K}{2 \pi \sqrt{N}} \sum_{j=1}^N \sin 2 \pi (q_i -q_j)\\  
q_i' &= q_i +p_i' \ .
\end{split}
\eeq
The kicking strength has been renormalized with $\sqrt{N}$ so that 
the final K-S entropy will be extensive. In much of the previous research with this model this normalization has been used primarily so that with the same kicking strength  increasing number of particles produces the same order of change in momentum for each of the particles.  This follows as the sum over the sines will be of the order of $\sqrt{N}$ if the $q_i$ mimic a random distribution.  There is an additional closely related system, for which it is also useful to make a few brief comments.  It has $N$ particles on a ring, but only the nearest neighbors (NN) interact via the above forces.   For this NN model, the renormalization of the interaction strength by $\sqrt{N}$ is not helpful and omitted.  All remarks on this system are clearly preceded by a reference to the NN model.

For $K=0$ the system is integrable since all the elements of the set $\{p_i \}$ are constants of the motion.   For more than a few particles, as $|K|$ is changed away from zero even by small amounts it is found numerically that the system develops significant chaotic dynamics.  Even for values of $|K|$ on the order of, but less than unity, it is already extremely difficult to detect any remnants of stable dynamics.  For a given configuration of particles the Hessian is given by:
\beq
\begin{split}
(\V'')_{ii}&=-\f{K}{\sqrt{N} }\sum_{j \ne  i} \cos 2\pi (q_i -q_j)\\
(\V'')_{ij}&= \f{K}{\sqrt{N} }\cos 2 \pi (q_i-q_j), \;\; i \ne j. 
\end{split}
\eeq
and the mass matrix is always the identity matrix.
\begin{figure}
\includegraphics[width=4.3in]{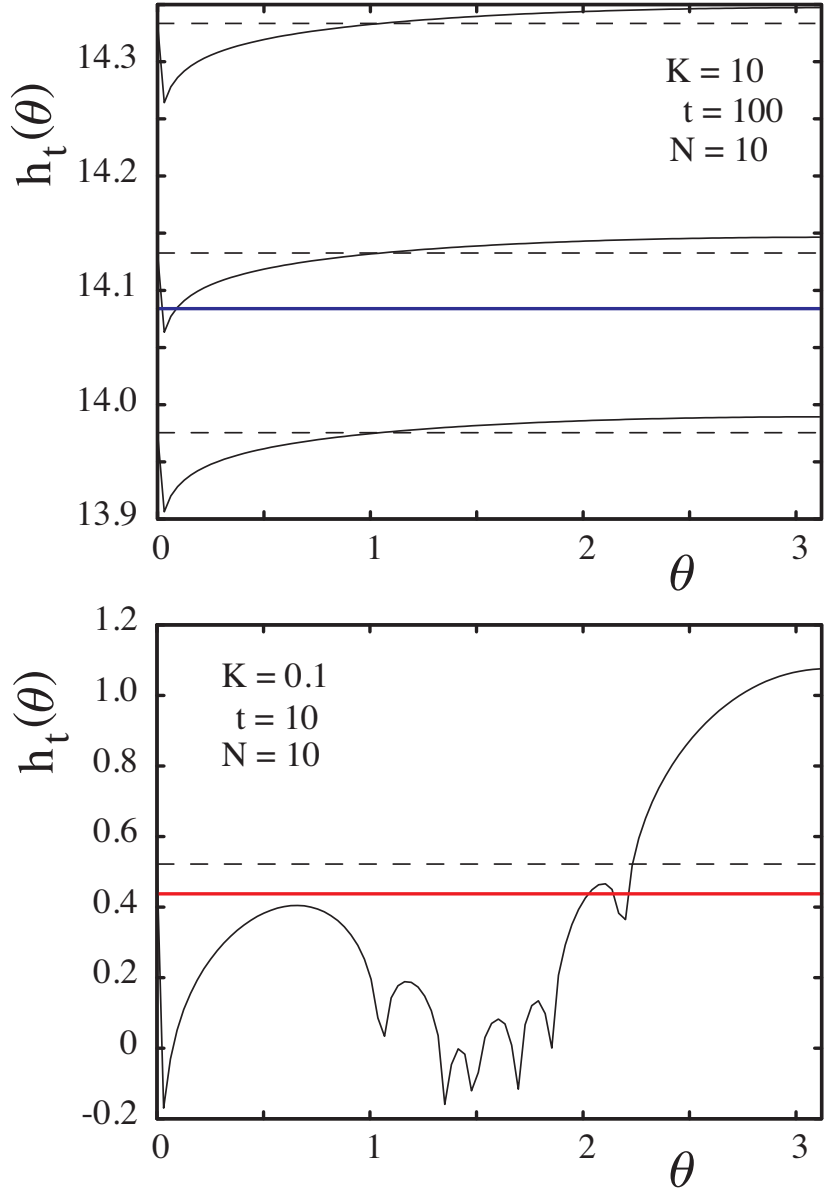}
\caption{The $\theta$ dependence of $h_t(\theta)$.  The upper panel shows $h_t(\theta)$ for three distinct initial conditions compared with the $\theta=0$ values shown as horizontal dashed lines; in each case, the known transient contribution $2\ln t/t$ is subtracted from $h_t(0)$.  All three $h_t(0)$ values here happen to match nearly exactly (at least 5 decimal place accuracy) the $\theta$ average.  The solid flat line is the average over initial conditions, thus giving some idea of the scale of the finite-time entropy fluctuations.  In the lower panel, a less well converged case, i.e. smaller $K$ and $t$, is shown.  There is more structure as a function of $\theta$ and the $\theta=0$ value is not equal in this case to the $\theta$-average shown here as the solid line.  Note that $h_t(\theta)$ has an exact reflection symmetry about $\pi$ and it is not necessary to show the range $(\pi,2\pi)$.\label{figure1}}
\end{figure}

The first step is to check the validity of averaging over $\theta$ in obtaining the finite time K-S entropy.  It is straightforward to show for $\theta=0$, i.e.~${\cal H}(1)$, that the center of mass coordinate generates a transient contribution to the finite time entropy of $2\ln t/t$ and this term is subtracted from $h_t (0)$ for the purposes of this discussion.  Figure \ref{figure1} illustrates the behavior of $h_t(\theta)$ in two regimes, one for stronger chaos and longer times and its opposite.  Remarkably, for each of the three initial conditions pictured in the upper panel, $h_t (0)$ equals the average of $h_t(\theta)$.  The $\theta$-dependence is essentially identical for each initial condition except for an overall displacement, and it is this displacement which contains all the information contained in the fluctuations of the finite-time K-S entropy.  For these parameters, the fluctuations are a small fraction of the infinite-time K-S entropy.  The lower panel shows the more complicated $\theta$-dependence found at shorter times and weaker chaos.  In this case, $h_t(0)$ does not equate to the $\theta$-average.  Curiously, after checking several other initial conditions, it appears as though the difference $h_t(0)-h_t$ only takes on a discrete set of values approximately.  Overall, it is found that replacing $h_t$ by $h_t(0)$ is an excellent approximation for large times.
\begin{figure}
\includegraphics[width=5in]{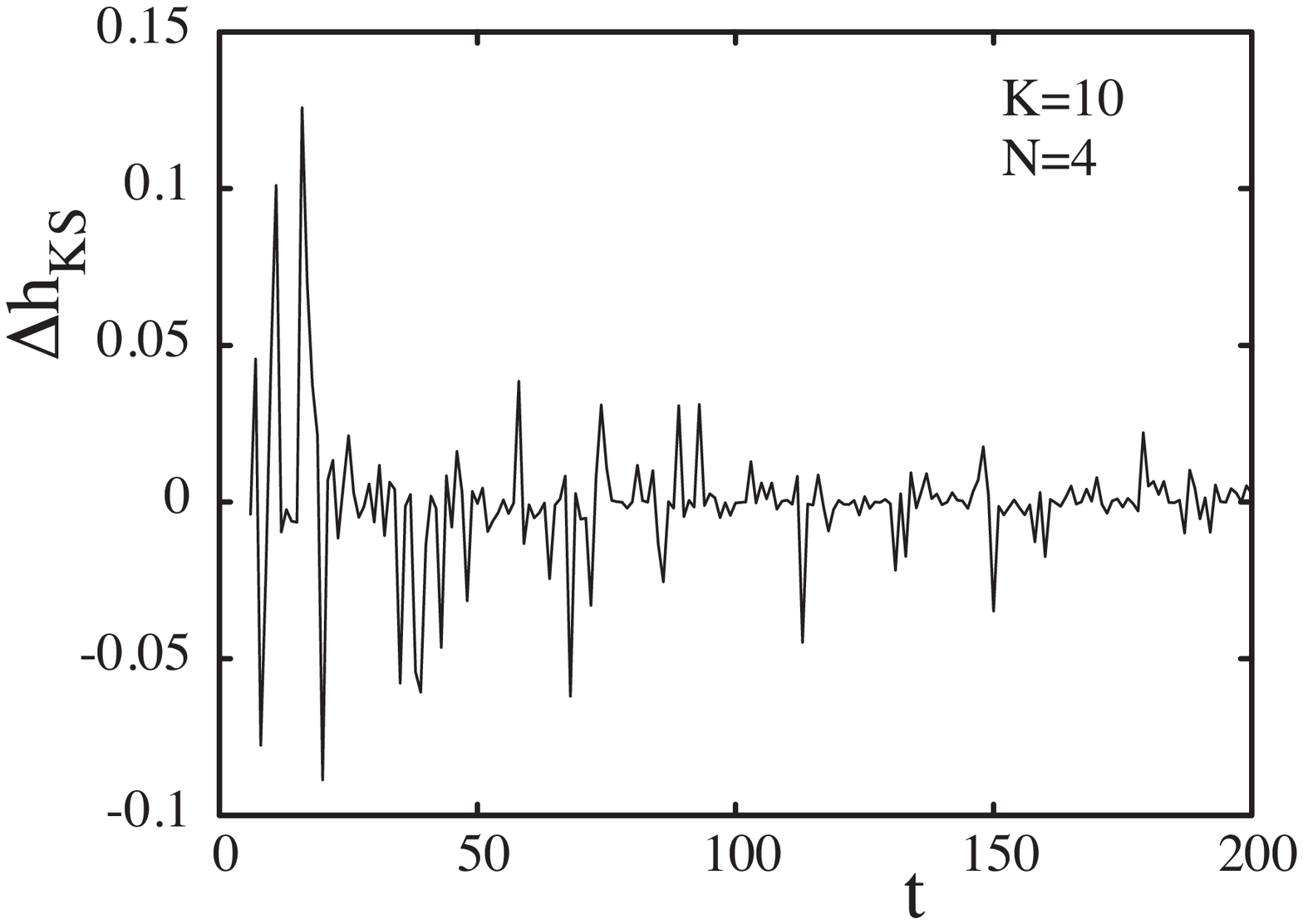}
\caption{The difference with time in the K-S entropy estimate calculated using the full ${\cal H}(1)$ and its banded counterpart in which the corner blocks are neglected.  The transient $2\ln t/t$ of ${\cal H}(1)$ at time $t$  and the transient $\ln(t+1)/t$ of the banded approximation were both subtracted before taking the difference between the two approximations.\label{figure2}}
\end{figure}

These results beg the question as to how well the banded approximation compares to $h_t(0)$.  Just as $h_t(0)$ has a center of mass transient term, so also has the banded matrix.  However, it is straightforward to show that its value is not the same as for $h_t(0)$.  Rather, it is $ \ln(t+1)/t$ and likewise is subtracted for the comparisons discussed here.  In Fig.~\ref{figure2}, the difference between the two approximations for $h_{KS}$ is plotted as a function of time for a reasonably strongly chaotic case.  There is no detectable difference not due to a random-like statistical error.  In fact, this statistical error is smaller than the finite-size fluctuations in $h_t$, and it may therefore be possible to get good estimates of the finite-time fluctuations even using the banded approximation, at least for strongly chaotic systems.  Though not shown here, the case $K=1$ had a very similar appearance except with a larger statistical error.

It remains only to understand its density of states under the ergodic assumption that 
the $q_i$ are uniformly distributed in $[0,1)$ over time and uncorrelated with respect to each other.  The density of states is not that of a full random symmetric matrix, namely the Wigner semicircle, as there are only $N$ random numbers input in determining the $N\times N$ Hessian matrix elements.  In addition, the diagonal elements are a sum of $N-1$ random contributions whereas the off-diagonal elements have just one.  Thus the spectra must reflect somehow this diagonal dominance.  

From Eq.~(\ref{diagapprox}), in the diagonal approximation finding this system's K-S entropy requires only the normalized spectrum of the potential's instantaneous Hessian.   In fact, the density of states has a very interesting and curious structure in which there are three exceptional eigenvalues, whereas the remaining eigenvalues form a Gaussian sea centered near zero. One of the three exceptional eigenvalues is the exact eigenvalue corresponding to the center of mass and the uniform mode, and the remaining two exceptional eigenvalues are ``outliers'', which for large particle number are well separated from the Gaussian sea formed by the others.

The outliers' origins may be seen by considering the matrix $A$, which results from the superposition of two one-dimensional projectors:
\beq
A=a( \nu_1 |\phi_1 \kt \br \phi_1 | + \nu_2 |\phi_2 \kt \br \phi_2 |),
\eeq
where $\br j |\phi_1 \kt = \cos 2\pi q_j/\sqrt{\nu_1}$, $\br j |\phi_2 \kt = \sin 2\pi q_j/\sqrt{\nu_2}$ and $\nu_1=\sum_j \cos^2 2 \pi q_j$,  $\nu_2=\sum_j \sin^2 2 \pi q_j$ ensure normalization. Then the element $A_{ij}$ is identical to the $i,j$ element of the Hessian $(\V'')_{ij}$ if the scaling factor $a=K/\sqrt{N}$.  Choosing this value of $a$, the diagonal elements of $A$ are constant and equal to $K/\sqrt{N}$.  As $A$ is the sum of  two one-dimensional projectors, it has exactly two non-zero eigenvalues, say $\lambda_1$ and $\lambda_2$, which can be found from $\tr(A)$ and $\tr(A^2)$.  The relations are:
\begin{equation}
\label{moments}
\lambda_1 + \lambda_2 = \tr(A)=N^{1/2}K\ ,\;\; \lambda_1^2 + \lambda_2^2 = \tr(A^2)= \frac{K^2(N+1)}{2} +\frac{K^2}{2N}\sum_{j,k=1 \atop j\ne k}^N \cos 4\pi (q_j-q_k) 
\end{equation}
whose solutions are
\beq
\lambda_{1,2} = \frac{KN^{1/2}}{2} \pm \frac{K}{2N^{1/2}} \sqrt{ \sum_{j,k=1}^N \cos 4\pi (q_j-q_k)}\ .
\eeq
These eigenvalues move toward infinity with increasing particle number and are each always close in value to $KN^{1/2}/2$.  In fact, their ensemble averaged root mean square displacement from this value is lower order and given by $K/2$.  Furthermore, 
it is possible to deduce a full probability density for these eigenvalues assuming the set $\{q_i\}$ behaves as uniformly random variables on the interval $[0,1)$, i.e.~the ergodic expectation for a fully chaotic dynamics.  One method is to calculate the moments,
\begin{equation}
\left<\left( \lambda_1 - \lambda_2 \right)^{2m} \right> = \left(\frac{K^2}{N}\right)^{m}\left< \left[ \sum_{j,k=1}^N \cos 4\pi (q_j-q_k) \right]^{m}\right> {\Longrightarrow \atop {N\rightarrow \infty}} K^{2m} m!
\end{equation}
which are the moments of an exponential probability density.   Another method is to realize that the sum is actually the sum of squares of two simple sums for which a central limit theorem applies to each and that leads to a $\chi_2^2$ probability density. Again, this is exponential and consistent with the moment method.  Therefore, for $t= \lambda_+ -\frac{KN^{1/2}}{2}$
\begin{equation}
\label{outlierrho}
\rho(t) = \frac{8t}{K^2}\exp\left(- \frac{4t^2}{K^2}  \right).
\end{equation}
Thus, the initial scaling of the parameter as $K/\sqrt{N}$ leads to the two nonzero eigenvalues of $A$ being of the order of $K\sqrt{N}/2$, and their fluctuations to be of order $K/2$ and hence independent of $N$.

The eigenvectors, $|\Psi_1\rangle$ and $|\Psi_2\rangle$, require a little bit more algebra to find because the two projector states are not orthogonal, i.e.~$\langle \phi_1|\phi_2\rangle \ne 0$.  However, it turns out that for a $\theta$ given just ahead, they can be expressed as
\beq
\begin{split}
\br j|\Psi_1\rangle &= \frac{\cos (2\pi q_j-\theta)}{S_1^{1/2}} \qquad \qquad S_1=\sum_{j=1}^N  \cos^2(2\pi q_j-\theta) \\
\br j|\Psi_2\rangle &= \frac{\sin (2\pi q_j-\theta)}{S_2^{1/2}} \qquad \qquad S_2=\sum_{j=1}^N  \sin^2(2\pi q_j-\theta),
\end{split}
\eeq
The determination of $\theta$ follows from imposing the conditions $\langle \Psi_2 |\Psi_1\rangle=0$ and $\langle \Psi_2 | A | \Psi_1\rangle=0$, which require that
\beq
0 = \sum_{j=1}^N\sin \left(4\pi q_j-2\theta  \right) \ .
\eeq
Denoting
\beq
s= \sum_{j=1}^N\sin \left(4\pi q_j  \right)\ {\rm and}\ t= \sum_{j=1}^N\cos \left(4\pi q_j  \right)\ {\rm gives}\ \theta = \frac{1}{2} \tan^{-1}\frac{s}{t}\ .
\eeq
Applying $A$ to the eigenvectors generates a relation between the value of $\theta$ and the eigenvalue difference, 
\beq
\frac{N}{K^2}\left( \lambda_1 - \lambda_2 \right)^2 = \left[ \sum_{j,k=1}^N \cos 4\pi (q_j-q_k)  \right] = \left( \sum_{j=1}^N \cos [4\pi q_j-2\theta]  \right)^2 
\eeq
or
\begin{equation}
2\theta = \phi - \cos^{-1}\left[ \frac{\pm N^{1/2}(\lambda_1-\lambda_2)}{K(\sqrt{s^2+t^2})} \right]  \qquad \qquad \phi = \cos^{-1} \frac{t}{\sqrt{s^2+t^2}}\ .
\end{equation}

Although the matrix $A$ is not the instantaneous Hessian, $2 I - \V^{\prime\prime}$, that is needed for the KS entropy problem, much of its spectral structure described above survives.  The critical distinctions between the two matrices are entirely related to the diagonal elements as the off-diagonal elements are connected exclusively by a sign change.   Alternatively, one could write that $2 I - \V^{\prime\prime}=2I-A+B$, where $B$ is a diagonal matrix whose diagonal elements are given by
\beq
B_{ii} = \sum_{j=1}^N A_{ij}\ .
\eeq
Relative to the spectrum of $-\V^{\prime\prime}$, the identity operator introduces a trivial shift by two of the entire spectrum and can be accounted for at the very end.  The $B$ matrix, on the other hand, is responsible for ensuring the exact center of mass eigenvalue (zero) and eigenvector (constant coefficients $N^{-1/2}$).  Note that the center of mass can have nothing to do with the KS entropy.  However, unlike the two previous levels of approximation in which the transient was decreasing with increasing time, the center of mass eigenvalue here leads to a constant contribution independent of time.  Its eigenvalue must be excluded from the spectrum of $\rho_V(\mu)$.  $B$ also introduces three more effects: i) the two projector states cease being exact eigenstates, ii) they have non-vanishing overlaps with the center of mass eigenvector, and iii) the remaining  $N-3$ formerly zero eigenvalues all take on non-zero values.  

Nevertheless, adding the diagonal elements is found neither to shift the two special levels significantly nor alter the width of their variation.  An educated guess as to the approximate structure of the perturbed projector eigenstates would be that the overlap with the center of mass eigenstate should be subtracted, $\theta$ slightly altered to maintain the orthogonality, and the normalization constants recalculated, i.e.
\beq
\begin{split}
\br j|\Psi_1\rangle &= \frac{1}{S_1^{1/2}}\left[ \cos (2\pi q_j-\theta) -\frac{1}{N} \sum_{i=1}^N  \cos(2\pi q_i-\theta)\right] \\
\br j|\Psi_2\rangle &= \frac{1}{S_2^{1/2}}\left[ \sin (2\pi q_j-\theta) -\frac{1}{N} \sum_{i=1}^N  \sin(2\pi q_i-\theta)\right]\ ,
\end{split}
\eeq
where, for example, $\theta$ would satisfy the relation
\beq
\sum_{j=1}\sin(4\pi q_j-2\theta) = \frac{2}{N} \sum_{j=1}^N  \cos(2\pi q_j-\theta) \sum_{i=1}^N  \sin(2\pi q_i-\theta) \ne 0
\eeq
which would give a new rotation differing from the previous one by an $O(N^{-1/2})$ shift.  As the center of mass eigenvalue is zero, removing its small eigenstate contribution to the previous projector eigenstates has little to no effect on the eigenvalues.  A priori, one might have anticipated that the loss of the $K/N^{-1/2}$ term from the diagonal might have moved these levels, but in fact that does not happen.

That leaves the $N-3$ eigenvalues to be understood.  They are related to the introduction of the diagonal elements, which are of higher order than the off-diagonal elements, and those elements are mostly accounted for by the projector states.  It is reasonable to expect and it is found that the remaining eigenstates are rather localized in nature except for having to be orthogonal to the 3 special extended states.  The diagonal elements behave like Gaussian random numbers from a central limit theorem, and one can anticipate that the $N-3$ eigenvalues form a Gaussian spectral sea.  Thus, to summarize the spectrum of $V''$ has a center of mass zero eigenvector with a uniform mode, which must be removed from the spectrum for entropy calculations, two delocalized modes with large outlier eigenvalues of the order of $K\sqrt{N}/2$, and the remainder of the spectrum's $N-3$ eigenvalues are in a Gaussian sea.  

The Gaussian sea has only two parameters to be determined, its centroid and width.  Its  properties can be determined by the first and second moments of the instantaneous Hessian similarly as done in Eq.~(\ref{moments}), except assuming ergodicity in the dynamics, which leads to ensemble averaging.  Given that 
\begin{equation}
\tr\left( 2I - \V^{\prime\prime} \right) = 2N + \frac{K}{N^{1/2}} \sum_{j,k=1 \atop j\ne k}^N \cos 2\pi (q_j-q_k)
\end{equation}
and that the sum of the three special eigenvalues equals $6-KN^{1/2}$ (now fully accounting for the identity operator also),  the mean location of the remaining $N-3$ levels $\langle \mu \rangle$ is
\beq
\label{gaussiancenter}
\langle \mu \rangle = \frac{1}{N-3} \left\langle 2N + \frac{K}{N^{1/2}} \sum_{j,k=1 \atop j\ne k}^N \cos 2\pi (q_j-q_k) + KN^{1/2} -6 \right\rangle=  2 + \frac{KN^{1/2}}{N-3}.
\eeq
By the same method, it is possible to also calculate the variance of $\rho_V(\mu)$.  Consider 
\begin{equation}
\left \langle \tr\left[ ( 2 I - \V^{\prime\prime} )^2\right] \right \rangle = 4N + K^2 (N-1),
\end{equation}
and subtract the contribution to this quantity from the three special levels to get the contribution for the remaining $N-3$ levels to the mean square operator trace.  Dividing by $N-3$ and subtracting the square of $\langle \mu \rangle$ gives the variance as
\begin{equation}
\label{gaussianwidth}
\sigma^2 = \f{K^2}{2}-\f{K^2N}{(N-3)^2}\ .
\end{equation}
Clearly, the separation of a Gaussian sea and 3 special eigenvalues does not make sense if $N<7$ as the variance becomes negative, which is impossible.  For large $N$, the width of the Gaussian sea approaches $K/\sqrt{2}$, which is the same order as the width of the distribution of the two outliers, and is again independent of $N$ due to the scaling of the kicking strength.  The centroid of the Gaussian sea is slightly more complicated.  For fixed $K$ and increasing $N$, it approaches $2$.  On the other hand, for a fixed, sufficiently large $N$, as $K$ increases ($K>> N^{1/2}$), it approaches $KN^{1/2}/(N-3)$.  Nevertheless, in either case for reasonably large $N$, the Gaussian sea does not overlap with the distribution of the two outliers, which are further out. 

All the ingredients for an evaluation of the K-S entropy within the diagonal approximation, Eq.~(\ref{diagonal}), are in place.  The two outliers contribute:
 \beq
 h_{KS}^o = \left \langle \ln\left|2 - \frac{KN^{1/2}}{2} -t \right| + \ln\left|2- \frac{KN^{1/2}}{2} + t\right| \right\rangle =  \int \ln\left|\left(2-\frac{KN^{1/2}}{2}\right)^2-t^2 \right| \rho(t) dt
 \eeq
 where $\rho(t)$ is given by the density in Eq.~(\ref{outlierrho}).  This integral can be evaluated exactly, 
 \beq
 h_{KS}^o= 2 \ln\frac{|K|}{2}+\ln\left(\frac{4}{K} - N^{1/2}\right)^2 - \exp\left[-\left(\frac{4}{K}-N^{1/2}\right)^2\right] \mbox{Ei}\left[\left(\frac{4}{K}-N^{1/2}\right)^2 \right]
 \eeq
 where $\mbox{Ei}(x)$ is the exponential integral.  The contribution from the Gaussian sea follows from a slight modification of Eq.~(\ref{gaussianentropy}) and  the final expression for the K-S entropy is 
 \beq
 \label{grandfinale}
  h_{KS}= h_{KS}^o + \left\{\begin{array}{l} (N-3)\left[\ln \sigma -{\gamma + \ln 2 \over 2} - \sum_{k=1}^{\infty}  \f{(-1)^k}{2k (2k-1)!!} \left( \f{\br \mu\kt}{\sigma}\right)^{2k} \right] \\   \\
(N-3)\left[ \ln \br \mu \kt - \sum_{k=1} \frac{(2k-1)!! }{2k} \left(\frac{\sigma}{\br \mu \kt } \right)^{2k}\right] \qquad \br \mu \kt >> \sigma  \end{array}\right.
\eeq
where $\sigma$ is the width given by Eq.~(\ref{gaussianwidth}) and $\br \mu\kt $ is the center of the Gaussian given in Eq.~(\ref{gaussiancenter}).  As in Eq.~(\ref{gaussianentropy}), the upper form is the convergent series and the lower form is the asymptotic series.
\begin{figure}
\includegraphics[width=5in]{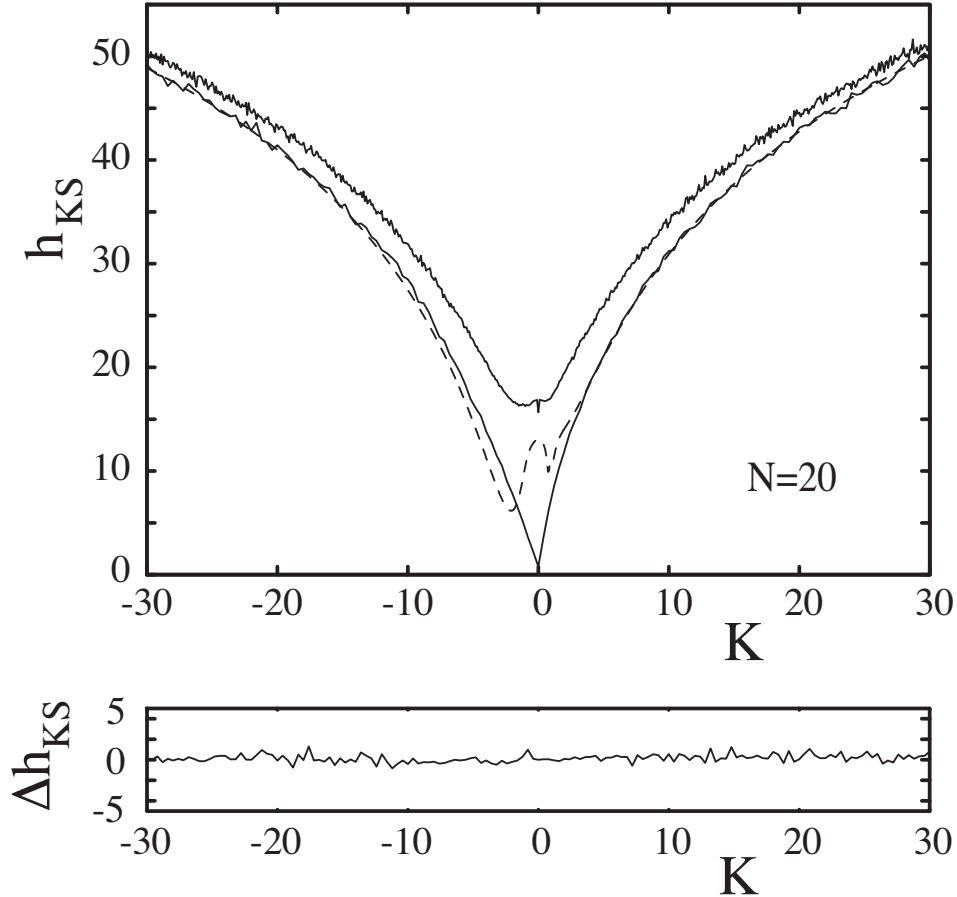}
\caption{The diagonal approximation of $h_{KS}$.  In the upper panel, the upper bound (highest lying curve) and the analytical form for the diagonal approximation (dashed curve) are compared to the banded approximation using $t=100$ and ten initial conditions.  In the lower rectangular box, the difference between the numerical diagonal approximation and the analytical form is shown.
\label{figure3}}
\end{figure}

The comparison of the upper bound, numerical diagonal approximation, and analytical diagonal approximation versus the banded approximation, which is taken to be essentially exact, as a function of interaction strength is shown in Fig.~\ref{figure3}.  It is seen that the upper bound does indeed lie above the other curves, and also as $K$ increases it gets closer and closer to the true K-S entropy.  Nevertheless, it is nowhere as good as the diagonal approximation.  The diagonal approximation is equally good as an analytical approximation or numerically evaluated approximation independent of the strength of the chaos, which is shown in the lower panel of the figure.  It has a pair of unusual double-well-like variations in its behavior for weakly chaotic systems (one of the double well-like variations is on too small a scale to show up clearly in the figure).  In this regime (small $K$), it is unable to follow the true K-S entropy.  The K-S entropy is seen to be an asymmetric function of interaction strength.  
The diagonal approximation does a bit better on the positive $K$ side.  It is impressive that the diagonal approximation works on an absolute scale.  On an entropy per particle basis, it would have even greater accuracy.  Overall, for moderate $K$ and $N$ the formula with all the details is necessary for the quality of the agreement with the K-S entropy shown.  

Note that it is well-known that the usual standard map (with one-degree of freedom) has a symmetry with respect to $K \leftrightarrow -K$ due to their phase spaces being shifts of each other by one-half along the position direction. For the fully interacting maps, this symmetry is broken.  This is seen in the $K$-dependence of the eigenvalues of the two special outlier states, which changes sign with the $K$ sign change (although the shift by two from the identity matrix does not).  Likewise, the off-centered density of the remaining $N-3$ states changes sign as well.  The NN model retains the symmetry exactly for even numbers of particles on the ring.  This can be demonstrated by shifting every other particle by one-half.  We have not identified a similar transformation for an odd number of particles, but confirmed numerically that the symmetry remains.

The entropy per particle for large $N$ and $|K|$ becomes [neglecting $O(K^{-4})$ and $O(N^{-1})$ contributions]
  \beq
  \f{h_{KS}}{N} \approx \ln \f{|K|}{2} -\f{\gamma}{2} + \f{4}{K^2}+\frac{4}{K\sqrt{N}}\ .
  \eeq
The leading term is the same as the one-degree-of-freedom kicked rotor noted in Eq.~(\ref{chirikov}), although the correspondence cannot be pushed to far because the meaning of $K$ is somewhat renormalized by removing the center of mass.  It is interesting to note that the leading term odd in the sign of $K$ goes a long way to explaining the difference in values of $h_{KS}$ for positive and negative $K$.  For example, from the last term in the above equation the difference of $h_{KS}$ at the edges of  Fig.~\ref{figure3} ($K= \pm 30$) is expected to be $1.192$, whereas it is actually $1.182$.
\begin{figure}
\includegraphics[width=4.5in]{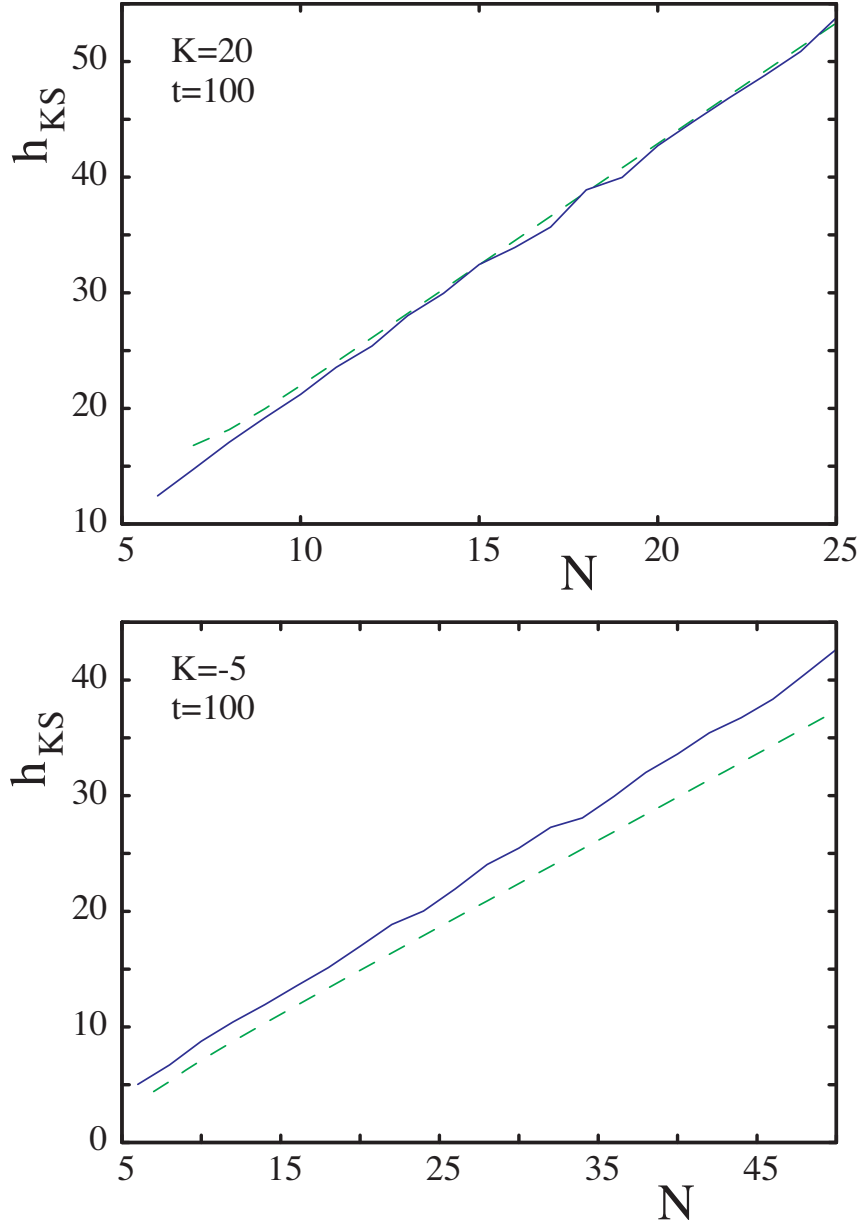}
\caption{The $N$-dependence of the analytical diagonal approximation.  The solid curves are the banded approximation using ten initial conditions.  The dashed curves are the analytical diagonal approximation.  \label{figure4}}
\end{figure}

Figure \ref{figure4} shows the diagonal approximation with fixed interaction strength and varying the number of particles.  For values of $K\lesssim-15$ or $K\gtrsim 5$, the diagonal approximation works extremely well.  In these $K$ regimes, above some value of $N$, the diagonal approximation matches the banded approximation to within sample size fluctuations (ten initial conditions were used).  In the upper panel, a $K$-value is shown where the approximation is working well.  As $N$ increases the diagonal approximation converges to the K-S entropy quickly with particle number.  The lower panel shows a weaker chaos case where the approximation is struggling a bit.  There the two curves slowly diverge.  For $-15<K<5$, increasing $N$ does not improve the estimate of the KS entropy in an absolute sense, although it would still improve on an entropy per particle basis.
\begin{figure}
\includegraphics[width=5in]{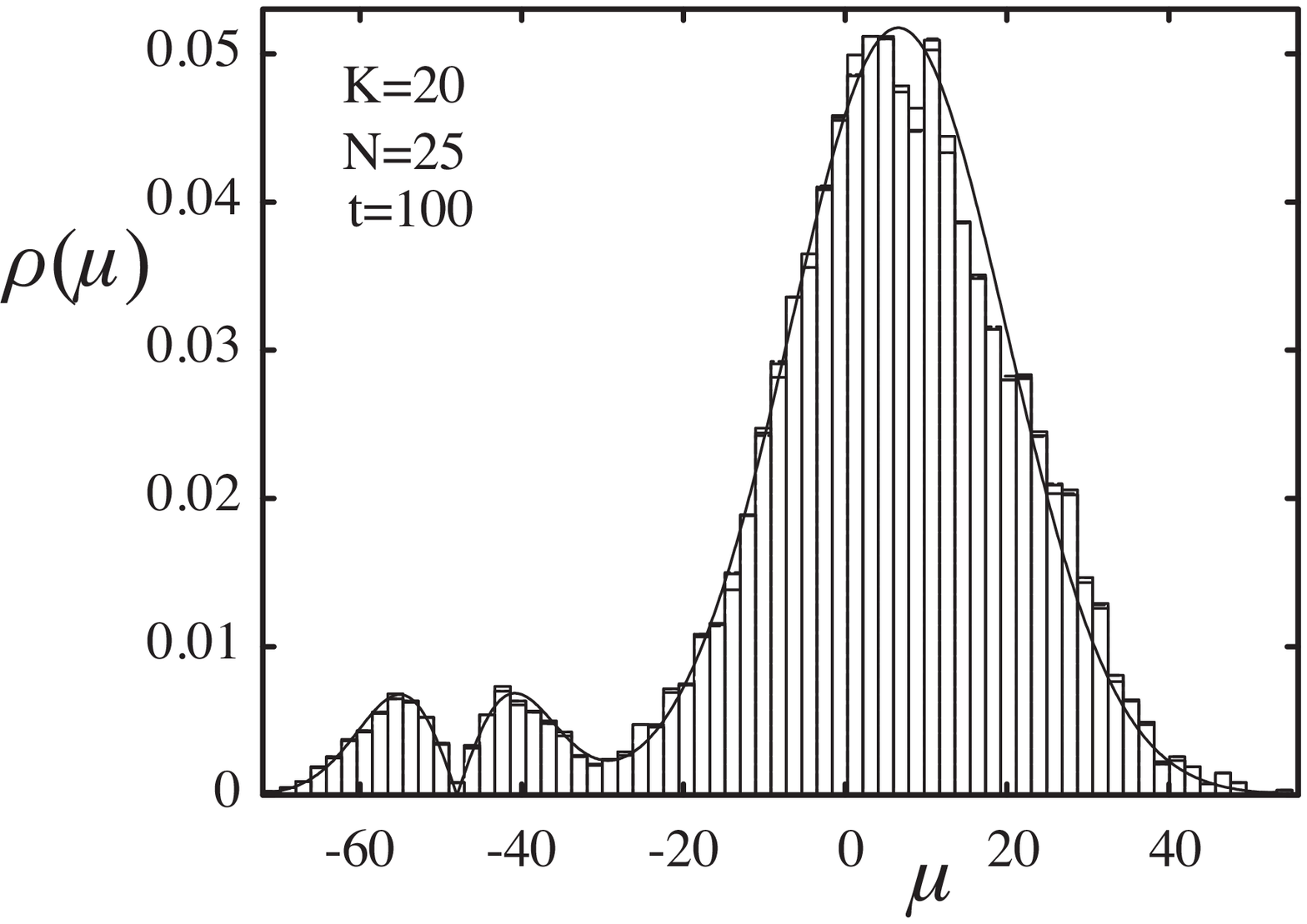}
\caption{The full, banded, and diagonal eigenvalue densities compared to the analytical form of the diagonal approximation.  The three histograms result from using $5$ initial conditions.  It is essentially impossible to see the difference between the full and banded results.   Small differences can be seen with respect to the diagonal case.
\label{figure5}}
\end{figure}

Finally, in Fig.~\ref{figure5} the eigenvalue densities are shown as histograms for the full ($h_t(0)$, banded, and diagonal approximations, and compared to the analytical expression for the instantaneous Hessians (diagonal approximation).  Excellent agreement is seen throughout the spectrum, including the details of the two outliers in the negative tail.  In the case of the NN model, the density of states does not have outliers, and is symmetric about zero. The absence of outliers follows from the lack of a rank-2 projector structure for the off-diagonal part of the Hessian matrix.  The density of states in the bulk is also not normal, as the diagonal elements do not look like sums of many random variables.  Thus, while the mean field model has more features like the outliers, analytical approaches to the density of states and hence, the K-S entropy was possible, in contrast to the NN model which needs other special methods not considered here.

\section{Summary and conclusions}

This paper presents a formalism to calculate the K-S entropy of symplectic maps of arbitrary dimension.  Constructing a transfer matrix, and using a duality relation, a formally exact expression for the finite-time K-S entropy is given as an average over a phase in the [logarithm of a] determinant of an effective Hamiltonian. The Hamiltonian has a tridiagonal block-banded structure with corner blocks which alone include the phase.  A series of approximations naturally follow in this scheme.  First, the variation with the phase is neglected and the resultant K-S entropy is found to  get better with longer time.  Second, the corner blocks of the Hamiltonian are themselves neglected resulting in a banded approximation that is found to be a very good approximation that is useful numerically, works even with weakly chaotic systems, and may still be valid for studying finite time fluctuations. The third level of approximation involves dropping of all off-diagonal blocks in the Hamiltonian and constitutes the diagonal approximation. This is expected to hold only in regimes of strong chaos.  A fourth estimate also follows as an exact upper-bound to the finite-time K-S entropy.  The last two, namely the diagonal approximation and the upper-bound can be found from the spectra of the instantaneous Hessians of the potential and are therefore extremely easy to evaluate numerically, vastly simpler and numerically faster than even the banded approximation.

One of the long term motivations was to study fluctuations in the finite-time K-S entropy, and we believe that neglecting phase averaging, one can still study such fluctuations for moderately chaotic regimes.  More investigation is needed to find out under what circumstances phase averaging can be left undone.  Curiously, many initial conditions, though not all, led to a zero angle value, $h_{t(0)}$, exactly equal to the phase average (exact here meaning to the accuracy of our calculations).  It appeared that differences when they appeared took on only a discrete set of values.  Furthermore, the banded approximation also appears form a useful and powerful tool to study fluctuations numerically in a dynamically interesting range of parameter values.  The diagonal approximation seems highly unlikely to reflect these fluctuations faithfully, especially in regimes of weak to moderate chaos.

A thorough study of a kicked version of the Hamiltonian Mean Field model, or coupled standard maps,  has been undertaken wherein the above approximations have been tested in detail.  It has been possible to find analytically the spectral density of the instantaneous Hessians and hence provide an analytical expression for the K-S entropy within the diagonal approximation.  For strongly chaotic systems, the diagonal approximation is an excellent estimate of the K-S entropy itself.  There is no real difference between the numerical diagonal and the analytical diagonal approximations.  
For weakly chaotic systems, the diagonal approximation is not good, but this is natural.  It is more accurate than the upper bound in all dynamical regimes.  It has strange double-well-like oscillations that is not an artifact of the extra approximations going into getting an analytic expression, as it is there in the numerical diagonal results as well.  It is also interesting that the K-S entropy is not an even function of $K$.  It turns out that the details of the analytical expression cannot be neglected for moderate values of $N$, and $K$.  For increasing particle number, the absolute error in the diagonal approximation is negligible except where it slowly grows in the weak chaos regime.  Nevertheless, even there the per particle entropy would improve for most values of $K$.

The analytical results pertaining to the diagonal approximation were facilitated by deriving the density of states for the instantaneous Hessians, which had a peculiar structure of a Gaussian sea along with two large outliers apart from the center of mass mode.  The density of states histograms for the full and banded approximations are difficult to distinguish whereas the diagonal case is a bit further away.  All of them match the analytic prediction for the level density extremely well.  It is interesting that given the long existence of the Chirikov estimate for the Lyapunov exponent or the K-S entropy of the two-dimensional standard map, this paper has finally provided a generalization to a nontrivial many-body extension of the same.  In addition, it is curious that an analytic diagonal approximation appears to be more difficult for the NN model even though it has a $K\leftrightarrow -K$ symmetry and thus its instantaneous Hessian possesses a symmetric density of states, and no outliers in its spectrum.

The techniques developed herein have several natural generalizations, for instance to 
systems with dissipation as well as to continuous time Hamiltonian flows. Even as it stands, this scheme will provide a pathway into the maze that is high dimensional chaotic phase spaces by allowing one to access relatively easily the finite-time K-S entropy and its fluctuations. The analytical structure behind the finite-time K-S entropy is also intriguing and needs to be more fully investigated.

\acknowledgments

The authors are indebted to hospitality and financial support of the Max-Planck-Institut f\"ur Physik komplexer Systeme in Dresden, GE where the work began.  One of us (ST) gratefully acknowledges financial support from the US National Science Foundation grant number PHY-0855337.

\bibliography{classicalchaos,quantumchaos}

\end{document}